\documentclass[
 reprint,superscriptaddress,
 amsmath,amssymb,
 aps,
prb,
]{revtex4-2}
\usepackage{graphicx}
\usepackage{dcolumn}
\usepackage{bm}
\usepackage[normalem]{ulem} 
\DeclareTextCommand{\DJ}{OT1}{%
  \raisebox{-0.1ex}{\scalebox{0.75}[1.4]{--}}\kern-.4em D%
}
\usepackage{xcolor}
\usepackage{subfigure}

\begin{document}

\preprint{APS/123-QED}

\title{Comparative analysis of plasmon modes in layered Lindhard metals and strange metals}

\author{Niels de Vries}
\email{niels@illinois.edu}
\affiliation{
Materials Research Laboratory, University of Illinois, Urbana-Champaign, 104 S. Goodwin Ave., Urbana, IL, 61801, USA
}
\author{Jin Chen}
\affiliation{
Materials Research Laboratory, University of Illinois, Urbana-Champaign, 104 S. Goodwin Ave., Urbana, IL, 61801, USA
}
\author{Eric Hoglund}
\affiliation{
Center for Nanophase Materials Sciences, Oak Ridge National Laboratory, Oak Ridge, TN 37830, USA
}
\author{Xuefei Guo}
\affiliation{
Materials Research Laboratory, University of Illinois, Urbana-Champaign, 104 S. Goodwin Ave., Urbana, IL, 61801, USA
}
\author{Dipanjan Chaudhuri}
\affiliation{
Materials Research Laboratory, University of Illinois, Urbana-Champaign, 104 S. Goodwin Ave., Urbana, IL, 61801, USA
}
\author{Jordan Hachtel}
\affiliation{
Center for Nanophase Materials Sciences, Oak Ridge National Laboratory, Oak Ridge, TN 37830, USA
}
\author{Peter Abbamonte}
\email{abbamont@illinois.edu}
\affiliation{
Materials Research Laboratory, University of Illinois, Urbana-Champaign, 104 S. Goodwin Ave., Urbana, IL, 61801, USA
}

\date{\today}

\begin{abstract}
The enigmatic strange metal remains one of the central unsolved problems of 21st century science. Understanding this phase of matter requires knowledge of the momentum- and energy-resolved dynamic charge susceptibility, $\chi(q,\omega)$, especially at finite momentum. Inelastic electron scattering (EELS), performed in either transmission (T-EELS) or reflection (R-EELS) geometries, is a powerful probe of $\chi(q,\omega)$. For the prototypical strange metal Bi$_2$Sr$_2$CaCu$_2$O$_{8+x}$, T-EELS, R-EELS, and infrared (IR) spectroscopy agree at $q \sim 0$, all revealing a highly damped plasmon near 1 eV. At larger $q$, however, EELS results show unresolved discrepancies. Since IR data are highly reproducible, it is advantageous to use IR data to calculate what the expected EELS response should be at modest $q$. Building on prior R-EELS work [J. Chen \textit{et al.}, Phys. Rev. B. \textbf{109}, 045108 (2024)], we extend this approach to T-EELS for finite stacks of metallic layers, comparing a ``textbook” Lindhard metal to a strange metal. In the Lindhard case, the low-$q$ response is dominated by long-lived, standing wave plasmon modes arising from interlayer Coulomb coupling, with in-plane dispersions that resemble the well-known Fetter modes of layered metals. This behavior depends only on the geometry and the long-ranged nature of the Coulomb interaction, and is largely insensitive to layer details. At larger $q$, the response reflects the microscopic properties of individual layers. For the strange metal, calculations based on IR data predict a highly damped plasmon with weak dispersion and no distinct surface mode. While our results match IR and R-EELS at low $q$, they do not reproduce any published EELS spectra at large $q$, highlighting unresolved discrepancies that demand further experimental investigation.
\end{abstract}

\maketitle

\section{\label{sec:introduction}Introduction}
Strange metals represent a class of quantum materials first observed in the 1980s \cite{Bednorz1986}. Strong correlations govern the microscopic electronic physics in these materials, as is evident by the peculiar behavior of the resistivity, quasiparticle self-energy, (optical) conductivity, and nuclear relaxation rates \cite{Varma1989}. Strange metallicity is found in a wide variety of materials, suggesting this may be a universal phenomenon, rather than a property of a specific system \cite{Phillips2022a}. This view is further supported by recent evidence of conformal invariance in the strange metal Bi$_2$Sr$_2$CaCu$_2$O$_{8+x}$, which indicates that there is no intrinsic length- or energy scale, the temperature being the only experimentally relevant parameter \cite{Guo2024a}. It is unknown what microscopic physics underlies the strange metal phenomenon, posing a profound problem for modern physics \cite{Zaanen2019a,Phillips2022a}.

To date, there is no accepted microscopic theoretical description of the strange metal phase \cite{Patel2023}. In the late 1980s, several authors introduced the marginal Fermi liquid (MFL) phenomenology, which posited that the polarizability, $\Pi(q,\omega)$, is linear in frequency ($\omega$) for $\omega < T$, remains constant with respect to energy for $\omega > T$ up to a cutoff, and is independent of momentum, $q$ \cite{Varma1989}. 
This form of the polarizability is unphysical and lacks a microscopic justification. However it reproduces many key features of the strange metal, including the linear-in-$T$ resistivity, the scaling of the quasiparticle self-energy, power-law behavior of the optical conductivity, and non-Korringa behavior of the nuclear spin relaxation rate \cite{Varma1989}. That such an {\it ad hoc} phenomenological model captures so many experimental signatures motivates the importance of measuring the momentum-dependent polarizability experimentally. 

The most direct way to measure the momentum-dependent polarizability of a metal is inelastic electron scattering, also known as electron energy-loss spectroscopy (EELS) \cite{Abbamonte2024}. EELS measures the dynamic charge susceptibility, $\chi(q,\omega)$, which in a homogeneous system is related to the polarizability by
\begin{equation}
\chi(q,\omega) = \frac{\Pi(q,\omega)}{1 - V(q)\Pi(q,\omega)} .
\end{equation}
Understanding the strange metal therefore requires determining the exact form of $\chi(q,\omega)$ using EELS techniques. 

Precisely such measurements were performed in the late 1980s, aiming to investigate the charge response of the strange metal Bi$_2$Sr$_2$CaCu$_2$O$_{8+x}$ with a transmission-geometry EELS (T-EELS) experiment. 
These measurements revealed a pronounced, strongly dispersing plasmon excitation \cite{Nucker1989,Nucker1991} whose behavior was consistent with predictions from the random phase approximation (RPA), which accurately describes free-electron metals \cite{Abbamonte2024}. These findings were fundamentally incompatible with the MFL phenomenology.

Additional experiments were carried out, including infrared (IR) optics measurements \cite{Basov2005,Levallois2016}, which probe the polarizability in the optical limit ($q = 0$), as well as further T-EELS studies \cite{Wang1990,Nucker1991,Terauchi1995,Terauchi1999a} and resonant inelastic x-ray spectroscopy (RIXS) studies on other cuprates\cite{Hepting2018,Nag2020}. While all techniques consistently observed a plasmon mode with an energy around 1 eV at small momentum ($q < 0.1 $\AA$^{-1}$), T-EELS results at larger momenta showed notable discrepancies. Some measurements reported a well-defined plasmon \cite{Nucker1989,Nucker1991}, whereas others found that no such excitation was present \cite{Terauchi1995,Terauchi1999a}.

Recent reflection-geometry EELS (M-EELS) experiments observe the same plasmon at zero momentum ($q \sim 0$) as other techniques \cite{Chen2024}, but do not detect a plasmon at finite momentum \cite{Mitrano2018,Husain2019}. Instead, the measured charge response at finite $q$ is momentum-independent and constant in energy---features that more closely resemble a MFL continuum than a dispersive, RPA-compatible plasmon. 
While all experimental probes are consistent at $q \sim 0$ \cite{Chen2024}, a clear discrepancy emerges at finite momentum. Resolving this discrepancy has foundational implications for our understanding of the strange metal problem and could significantly advance our broader understanding of strongly interacting electron systems in general.

Reflection and transmission EELS techniques differ in several important ways. First, the two geometries probe distinct---though related---charge response functions \cite{Vig2017,Abbamonte2024}. Second, the sample requirements differ: R-EELS probes a single surface of a specimen, whereas T-EELS requires a sample thin enough for the electron beam to transmit through, so two surfaces are always probed. Because material surfaces often host characteristic excitations that differ from those of the bulk, relating both R-EELS and T-EELS data to the intrinsic bulk charge susceptibility can be subtle and requires careful analysis.

For homogeneous systems, the relationship between the experimental EELS cross-section and the bulk charge susceptibility has been thoroughly investigated for both transmission and reflection geometries \cite{Bethe1930,Ritchie1957,Batson1983,Sturm1993,Egerton2011,Evans1972,Ibach1982,Mkhoyan2007,Kogar2014,Vig2017}. However, this relationship is not as well established for layered or inhomogeneous systems. A recent study addressed this issue for R-EELS from a layered strange metal, showing good agreement with infrared (IR) optical measurements \cite{Chen2024}. An equivalent analysis for the transmission geometry, however, has not yet been done, and is essential for interpreting T-EELS measurements of layered materials.

Understanding the relationship between transmission and reflection EELS---specifically in how their respective experimental cross-sections relate to the bulk dynamic charge susceptibility---is essential for resolving discrepancies over the observed charge response of strange metals.
This leads to the aim of this work, which is to investigate the charge response of a (layered) strange metal of finite thickness, bounded by two surfaces, in a T-EELS experiment. Our approach is to derive the experimental T-EELS cross-section for a finite, layered system, assuming the sample is sufficiently thin for the Born approximation to hold, while fully incorporating boundary excitations that exist because of the finite size of the system. We analyze which features of the response arise generically from the layered geometry, and which derive from the microscopic properties of the constituent layers.

We compare two cases. In the first, the T-EELS cross section of a layered electron gas (LEG) is computed using Lindhard theory within the random phase approximation (RPA), assuming the layers are coupled only by the direct Coulomb interaction. 
In the second, we calculate the same cross section for a layered strange metal, using the individual layer polarizability extracted from IR optics data at $q \sim 0$, which is highly reproducible across multiple experiments \cite{Basov2005,Levallois2016,Marel2003b,Chen2024}.
This comparison will reveal what T-EELS results are expected based on what is known from IR experiments, and also reveal 
which response features are generic to layered metals and which are specific to the strange metal state, at least at modest values of $q$. 

\section{T-EELS from a layered material}
T-EELS is a well-suited probe for measuring the dynamic charge response of materials. The charge response represents the probability that a perturbation in the charge density at position and time $\vec{r}^{\prime},t^{\prime}=0$ propagates to $\vec{r},t$, and is written in terms of the bosonic density operator $\hat{\rho}(r,t)$:
\begin{equation}
\label{eq:densityresponse}
\chi(r,r^{\prime},t,0) = - \frac{i}{\hbar} \left<[\hat{\rho}(r,t),\hat{\rho}(r^{\prime},t^{\prime})] \right> \Theta(t-t^{\prime})
\end{equation}
where $\left<x\right>$ represents the expectation value of $x$, $[,]$ is a commutator, and $\Theta(\tau)$ is a Heaviside function, which guarantees causality. In a homogeneous, infinite, three-dimensional system, $\chi(\vec{r},\vec{r}^{\prime},t) = \chi(\vec{r}-\vec{r}^{\prime},t)$ and is usually given as its Fourier transform $\chi(\vec{Q},\omega)$. The latter quantity is commonly related to the T-EELS cross-section\cite{Egerton2011,Nicholls2019} as follows:
\begin{equation}
\frac{d^2\sigma}{d\Omega d\omega} = - \frac{m^2 e^2}{\hbar^3 |Q|^4 \epsilon_0^2 4 \pi^3}\frac{k_1}{k_0}\frac{1}{N_s}\frac{1}{1-e^{-\hbar\omega / k_BT}}\chi^{\prime\prime}(\vec{Q},\omega),
\end{equation}
in which $N_s$ is the number of scatterers, $\vec{Q}$ is the momentum transfer, $T$ the temperature, $\chi^{\prime\prime}$ denotes the imaginary part of $\chi$, and $k_0$ and $k_1$ are the wavenumbers of the incident and scattered probe electron, respectively.

In a layered system, translational symmetry is broken in the $\hat{z}$-direction, and $\chi$ is now an explicit function of $z$ and $z^{\prime}$. It is therefore necessary to specify the T-EELS cross-section in mixed form, $\chi(\vec{q},z,z^{\prime})$, which is related to $\chi(Q,\omega)$ as follows:
\begin{multline}
\label{eq:chi_inplaneq}
\chi(\vec{Q},\omega) = \chi(\vec{Q},-\vec{Q},\omega) = \chi(\vec{q},q_z,-q_z,\omega)\\
= \int \mathrm{d} z \mathrm{d} z^{\prime} \chi(\vec{q},z,z^{\prime},\omega) e^{-i q_z z} e^{i q_z z^{\prime}},
\end{multline}
in which $\vec{Q}$ is now split into its component along the layer planes, $\vec{q}$, and the perpendicular component, $q_z$. This gives the more general form for the T-EELS scattering cross-section:
\begin{multline}
\label{eq:layeredcrosssection}
\frac{d^2\sigma}{d\Omega d\omega} = \frac{m^2 e^2}{\hbar^3 |Q|^4 \epsilon_0^2 4 \pi^3}\frac{k_i}{k_s}\frac{1}{N_s}\frac{1}{1 - e^{-\hbar \omega / k_B T}} \\
\times \mathrm{Im}\left[\int \mathrm{d} z \mathrm{d} z^{\prime} \chi(\vec{q},z,z^{\prime},\omega) e^{i q_z (z^{\prime} - z)}\right].
\end{multline}

This equation applies to any layered material with arbitrary stacking. The goal of this work is to understand the T-EELS response of a layered, finite system, and so we need to consider the quantity $\chi(\vec{q},z,z^{\prime},\omega)$ for such a system. 

A key difference between the current case and that of an infinite, layered system is that the latter lacks distinct surface excitations, and a closed-form relationship between $\chi(q,\omega)$ and the polarizability, $\Pi(q,\omega)$, is well known \cite{Fetter1973,Fetter1974,Shung1986,Reed2010}. In the semi-infinite case, a closed form expression for $\chi(\vec{q},z,z^{\prime},\omega)$ was obtained by Jain and Allen \cite{Jain1985}, who showed that the broken translational symmetry gives rise to distinct excitations characteristic of the surface. 

For a finite-sized, layered system, such a closed-form expression does not exist. It is therefore necessary to carry out explicit numerics using equation \ref{eq:layeredcrosssection} for some specific cases to understand the exact behavior of bulk and boundary modes.

\section{Density response for a finite, layered metal}
\label{sec:framework}
In order to compute $\chi(\vec{q},z,z^{\prime},\omega)$ for a finite-sized, layered system, we adapt the approach for a semi-infinite system derived by Jain and Allen \cite{Jain1985}. Their original derivation expresses $\chi(\vec{q},z,z^{\prime},\omega)$ analytically, for a layered electron system with a single surface, in terms of the 2D polarizability of a single layer, $\Pi^0(\vec{q},\omega)$. It was used to compute the Raman response of a GaAs-AlGaAs layered heterostructure, but the EELS response can be derived from it as well \cite{Chen2024}. In this section, we define the geometry and introduce the framework we use to compute $\chi(\vec{q},z,z^{\prime},\omega)$ for a finite-sized, layered system.

\begin{figure}
\includegraphics[width=0.6\columnwidth]{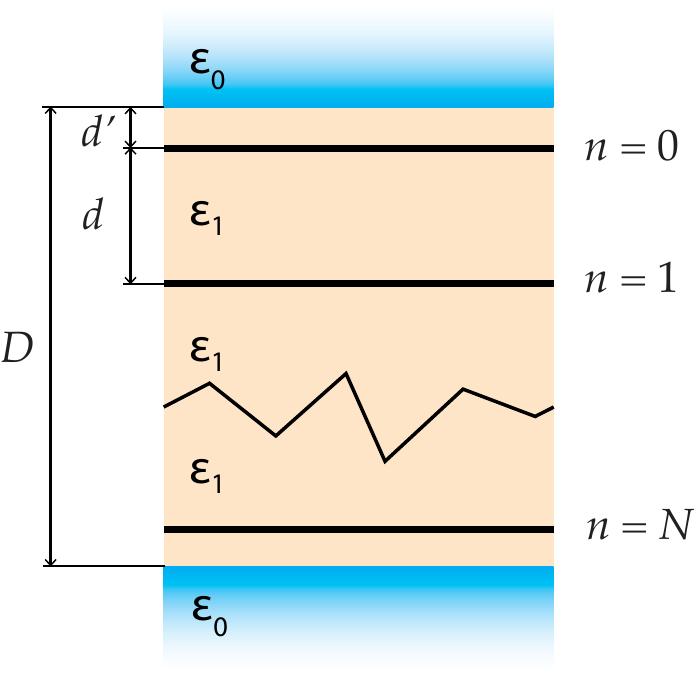}
\caption{ The geometry of the model of $N$ layered electron gas planes. $D$ is the total thickness of the structure, $d$ the interlayer distance, and $d^{\prime}$ the distance between the outer layers and the vacuum-dielectric interface.
\label{fig:geometry} }
\end{figure}

Analogous to Jain and Allen, we consider a layered structure consisting of $N$ conducting planes that are homogeneous in the $\hat{x}$- and $\hat{y}$-directions and discrete in the $\hat{z}$-direction. These layers are assumed to be very thin,
and sufficiently far apart that interlayer tunneling may be neglected, so that the layers interact only through the long-ranged Coulomb interaction \cite{Jain1985}. 
As the system is isotropic and homogeneous in the $\hat{x},\hat{y}$-directions, the in-plane momentum $\vec{q}$ can be written as a scalar $q$. The charge density $\rho(x,y,z) = \delta(z-nd)$ in which $n \in \{ 1, 2, ... N \} $, $d$ being the interlayer distance. The space between the layers is treated as a homogeneous dielectric with dielectric constant $\epsilon_1$. At distances $-d^{\prime}$ and $d^{\prime}$ from the termination layers at $n=0$ and $n=N$, respectively, there is an interface with the vacuum, as illustrated in Fig. \ref{fig:geometry}. Each individual  layer is assumed to have the same 2D polarizability, $\Pi^0(q,\omega)$.

Computing $\chi(\vec{q},z,z^{\prime},\omega)$ for a finite-thickness, layered system differs from the approach of Refs. \cite{Jain1985,Chen2024} in two important respects. The first derives from the existence of a second surface. In the semi-infinite system, there is only one surface, which is modeled electrostatically using an image charge correction. In the finite system there are two surfaces, requiring a large number of image charges, since the two surfaces create images that are reflected in one another as an infinite series. Second, while a closed-form, analytic solution was obtained for the semi-infinite case \cite{Jain1985}, no such relation is known for the finite case. We are therefore restricted to numeric solutions, as described in further detail below. 

\subsection{Image charge corrections for a finite slab}
\label{sec:image_charge_corrections}
The Coulomb interaction between charges situated in different layers is defined as 
\begin{equation}
V(\vec{r},\vec{r}^{\prime}) = \frac{e^2}{4 \pi \epsilon_0 \epsilon |\vec{r} -\vec{r}^{\prime}|} + V_{\mathrm{im}}(\vec{r},\vec{r}^{\prime}),
\end{equation}
where $V_{\mathrm{im}}$ represents the correction to the Coulomb interaction due to the interfaces with the vacuum. The dielectric response of the system is discontinuous at the two surfaces, which can be corrected using an image charge. However, as there are two surfaces, image charges originating from the correction of one surface need to be corrected at the other surface, which leads to an infinite number of image charge corrections. Higher order surface charge corrections contribute less due to the Coulomb potential decaying with distance. This means that the corrections can be recursively computed until the error is within a certain tolerance, as additional terms become small. This gives an expression for $V_{\mathrm{im}}(\vec{r},\vec{r}^{\prime})$. Analogous to Eq. \ref{eq:chi_inplaneq}, the mixed Fourier transform $V(q,z,z^{\prime})$ has the form:

\begin{equation}
V(q,z,z^{\prime}) = \frac{e^2}{4 \pi \epsilon_0 \epsilon q} \left[ e^{-q |z-z^{\prime}|} + f_{\mathrm{im}}(q,z,z^{\prime}) \right],
\end{equation}
in which $f_{\mathrm{im}}(q,z,z^{\prime})$ is a form factor that describes the image charge corrections.

In order to compute $f_\mathrm{im}$, it is convenient to define a parameter $D = (N-1)d + 2d^{\prime}$ which is the total thickness of the structure, i.e. the distance between the two interfaces. Analogous to the Jain-Allen approach, we define the charge of the $m$\textsuperscript{th} image charge to be $\alpha^{m+1}e$, where $\alpha = (\epsilon_1 -1)/(\epsilon_1 + 1)$. Using a number of parity functions, the $z^{\prime}$-position of the image charge can be expressed for arbitrary $m$. As both interfaces of the slab required image charge corrections, the function $f_{\mathrm{im}}(q,z,z^{\prime})$ consists of two series. These series always converge as $|z-z^{\prime}|$ increases with $m$, but they convergence faster if $\alpha < 1$. The image charge contribution is written as $f_{\mathrm{im}}(q,z,z^{\prime}) = \sum_{m=0}^{\infty}\left(b_m(q) + c_m(q)\right)$ with $b_m(q)$ and $c_m(q)$ being functions describing the image charge recursions starting at the $n=0$ and $n=N$ surfaces, respectively. They are defined as
\begin{equation}
\begin{cases}
b_n(q,z,z^{\prime}) = \alpha^{n+1} e^{-q \left|z - \zeta(n) \left( z^{\prime} + \xi(n) \cdot D + \eta(n) \cdot 2d^{\prime} \right) \right|} \\
c_n(q,z,z^{\prime}) = \alpha^{n+1} e^{-q \left|z - \zeta(n) \left( z^{\prime} - \xi(n+1) \cdot D + \eta(n) \cdot 2d^{\prime} \right) \right|}
\end{cases}
\end{equation}
using the parity functions
\begin{equation}
\begin{cases}
\zeta(x) &= 2 (x\mod 2)-1, \hspace{5mm} (-1, 1, -1, 1, -1, ...)\\
\xi(x) &= x + (x\mod 2), \hspace{5mm} (0, 2, 2, 4, 4, ...)\\
\eta(x) &= (x+1)\mod 2. \hspace{5mm} (1, 0, 1, 0, 1, ...)
\end{cases}
\end{equation}

\subsection{Calculation of the charge response}
\label{sec:computation}
Compute the inelastic scattering cross-section requires determining the dynamic charge susceptibility $\chi(z,z^{\prime},q,\omega)$ in the presence of Coulomb interactions. This quantity may be expressed as a Dyson equation: 
\begin{multline}
\label{eq:dyson}
\chi(z,z^{\prime},q,\omega) = \Pi^0(q,\omega) \delta(z,z^{\prime}) \\ + \Pi^0(q,\omega) V_{2D}(q) \sum_{\tilde{z}}f(q,z,\tilde{z}) \chi(\tilde{z},z^{\prime},q,\omega) . 
\end{multline}
Here, $V_{2D}(q) = e^2/4\pi\epsilon_0q$ is the two-dimensional Fourier transform of the in-plane Coulomb interaction. Inverting this implicitly gives
\begin{equation}
\label{eq:inv_dyson}
\Pi^0(q,\omega) \chi^{-1}(z,z^{\prime},q,\omega) = \delta(z,z^{\prime}) - \Pi^0(q,\omega)V_{2D}(q) f(q,z,z^{\prime}).
\end{equation}
There being no known analytic solution for a finite system, we take the approach of computing $\chi(z,z^{\prime},q,\omega)$ numerically, which requires prior knowledge of $\Pi^0(q,\omega)$ either either from a theory or a measurement. 

Because the quantities $z,z^{\prime}$ are discrete layer indices, the response $\chi(z,z^{\prime})$  can be expressed as an $N \times N$ matrix. We define the right hand side of Eq. 11 as $\mathbf{B}^{-1}$, which allows us to express the problem, at fixed $q,\omega$, as a linear matrix equation:
\begin{equation}
\label{eq:numerical_dyson}
\boldsymbol{\chi} = \Pi^0 \mathbf{B}, \\
\end{equation}
\begin{equation}
\label{eq:numerical_B}
\mathbf{B} = \left( \mathbf{I} - \Pi^0 V \mathbf{F}\right)^{-1},
\end{equation}
in which $\boldsymbol{\chi}$ and $\mathbf{F}$ are matrices defined as $\chi(z,z^{\prime})$ and $f(q,z,z^{\prime})$ respectively. $\mathbf{I}$ is the identity matrix, and $\Pi^0$ and $V$ are scalars defined as $\Pi^0(q,\omega)$ and $V_{2D}(q)$, respectively.

To calculate $\chi(z,z^{\prime})$ at given $q,\omega$, we first define the system with the number of layers, the layer spacing, and the single layer polarizability $\Pi^0(q,\omega)$. The Coulomb interaction potential $V(\vec{r},\vec{r}^{\prime})$ is then computed as discussed in section \ref{sec:image_charge_corrections}. Then, $\mathbf{B}$ is computed from Eq.  \ref{eq:numerical_B} by numerically inverting using a lower-upper decomposition scheme. $\chi(q,z,z^{\prime},\omega)$ is then computed using Eq. \ref{eq:numerical_dyson}, and the T-EELS cross-section is subsequently computed using Eq. \ref{eq:layeredcrosssection}.

\section{Reponse of a layered Lindhard electron gas}
Before investigating a strange metal, we first consider, for comparison, the T-EELS response of a simple, textbook, layered Lindhard electron gas in the random phase approximation (RPA), which assumes the quasiparticles have infinite lifetime \cite{Pines2018}. The layers are assumed to interact only through the direct Coulomb interaction. We aim  to understand what aspect of the response is generic to the Coulomb interaction and the layered geometry of the system, and what aspect explicitly depends on the form of the polarizability $\Pi^0(q,\omega)$, i.e., the microscopic properties of the individual layers. 
We therefore compute the full T-EELS cross section in two cases with the same geometry---and hence the same structure factor for the Coulomb interaction---but with different polarizabilities, $\Pi^0(q,\omega)$. Comparing the responses of these two cases will reveal what part of the dynamic response originates from the interlayer Coulomb interaction, and what part can be attributed to the microscopic layer properties. The two polarizabilities we consider are (1) the full, two-dimensional, frequency- and momentum-dependent Lindhard polarizability, and (2) its value in the limit of high frequency and small-momentum. 

In this section, we first introduce the functional form of the full 2D Lindhard polarizability, and then determine its form in the small-q limit. The T-EELS response is then calculated for these two cases. The results for the latter case are shown in section \ref{sub:small_q_limit_polarizability}, and the former case in Section \ref{sub:full_2d_lindhard_polarizability}. The T-EELS data for the two cases are then compared. The difference between the computed responses in the two cases will provide insight into what part of the T-EELS response can be attributed to the interlayer Coulomb interaction, and what part can be attributed to the microscopic polarizability.

\subsection{Lindhard polarizability in two dimensions}

The Lindhard polarizability of a two-dimensional electron gas was computed in RPA exactly decades ago \cite{Stern1967}, and takes the following form \cite{Mihaila2011}:
\footnotesize
\begin{equation}
\label{eq:lindhard}
\begin{cases}
\Pi^{0\prime}(q,|\omega|) = \frac{N_e}{E_F}\left[ -1 -C_{-} \frac{1}{q}\sqrt{\left|\frac{q^2_{-}}{2q}\right|^2 -1} +C_{+} \frac{1}{q]}\sqrt{\left|\frac{q^2_{+}}{2q}\right|^2 -1} \right]\\
\Pi^{0\prime\prime}(q,|\omega|) = -\frac{N_e}{E_F}\frac{1}{q}\left[ \sqrt{1 - \frac{1}{4}\left(\frac{\nu}{q} - q\right)^2} - D\sqrt{1-\frac{q^2}{4} -\frac{\nu}{2} - \frac{\nu^2}{4q^2}}\right] \\
\Pi^0(q,\omega) = \Pi^{0\prime}(q,|\omega|) + i\frac{\omega}{|\omega|}\Pi^{0\prime\prime}(q,|\omega|) .
\end{cases}
\end{equation}
\normalsize
Here, $N_e$ is a two-dimensional electron number density, $E_F$ is the Fermi energy, $q$ is the excitation momentum normalized to the Fermi wavevector, $k_F$, and $\nu = \hbar\omega/E_F$ is the normalized energy. $q_{\pm} \equiv \nu \pm q^2$, and $C_{\pm}$ and $D$ are parity functions defined as
\begin{equation}
D_{\pm} = \mathrm{sgn}(q_{\pm}^2) \Theta(|q_{\pm}^2|/(2q) -1)
\end{equation}
and
\begin{equation}
C = 
\begin{cases}
1 & (q < 2, 0 \le \nu < -q^2 + 2q) \\
0 & \nu \ge -q^2 + 2q . 
\end{cases}
\end{equation}

For $\nu \gg q^2 + 2q$, the imaginary part $\chi^{0\prime\prime}(q,\omega)$ vanishes, and the real part reduces to 
\begin{equation}
\label{eq:simplefermigas}
\chi^{0} = \frac{\pi N q^2}{m_{\mathrm{eff}} \omega^2}.
\end{equation}
This is the two-dimensional Lindhard polarability in the small $q$ limit. Note that this quantity is purely real, so the collective modes that result will experience no Landau damping. As this would result in a response taking the form of a series of Dirac delta-functions, a small but finite imaginary constant is added to artificially broaden the response. 

To make the comparison with strange metals as close as possible, we choose the layer spacing and carrier density to be the same as Bi-2212. The Fermi energy is set to 0.52~eV, and the two-dimensional electron number density $N = k_F^2/2\pi$ = $2.19 \cdot 10^{18}$~m$^{-2}$, which corresponds to a doping of $p=0.16$/Cu. The spacing between the CuO layers is chosen to be half the lattice constant, $d$ = $c/2$ = 15.4~\AA, and the background dielectric constant $\epsilon_1=4.5$ is taken from IR optics \cite{Levallois2016}. The parameter $d^{\prime} = 0$, meaning the top- and bottom-most layers are the interfaces with the vacuum.

\subsection{Small-q limit polarizability}
\label{sub:small_q_limit_polarizability}

\begin{figure*}
\includegraphics[width=\textwidth]{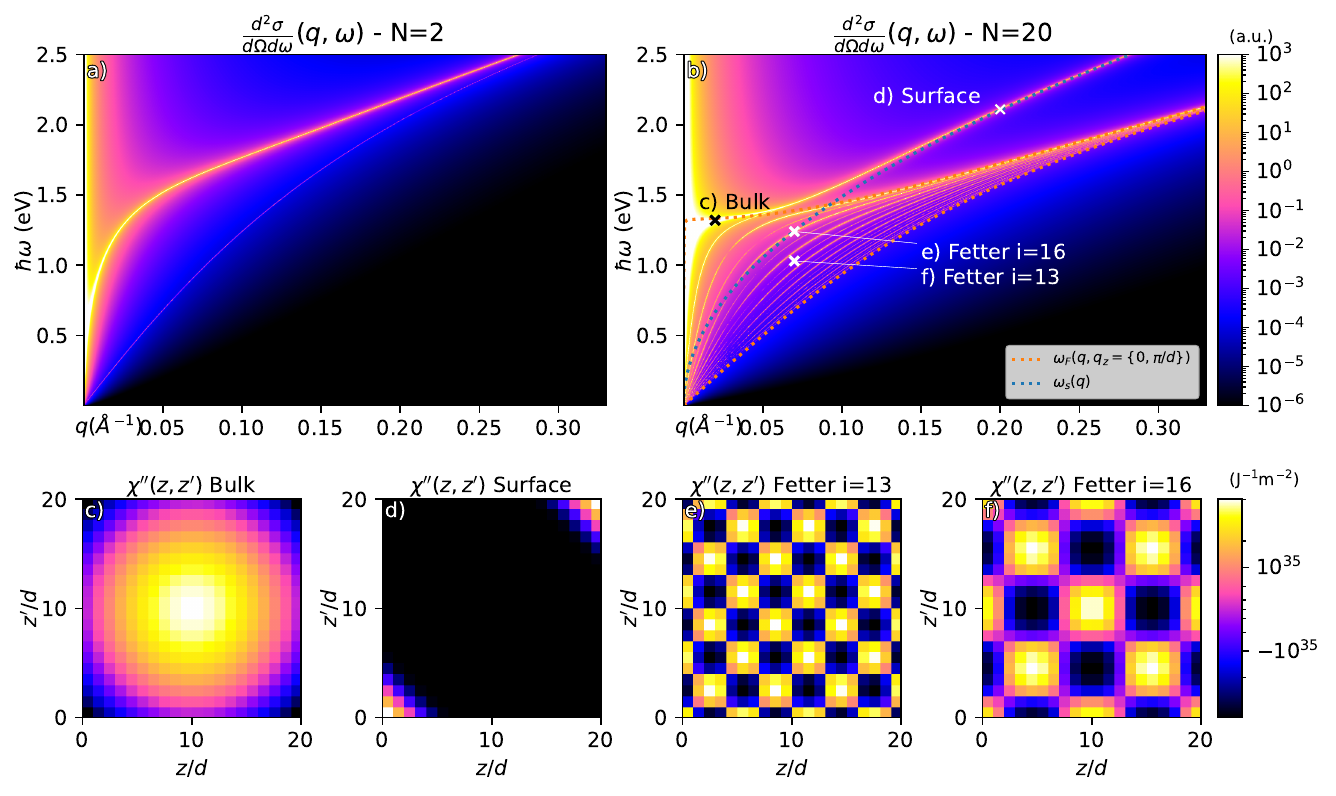}
\caption{The simulated transmission scattering cross-section of (a) an $N=2$ layer electron gas system plotted on a ($q,\omega$)-grid, and (b) an $N=20$ layer electron gas system plotted on a ($q,\omega$)-grid. For both calculations, the incident beam energy is taken as 60~keV. Due to momentum conservation, the momentum transfer $q_z$ parallel to the beam is $<0.01$~\AA$^{-1}$. Here, the dashed white line shows the dispersion of the surface plasmon given by Eq. \ref{eq:surfplasmon}. The dashed orange lines represent the $q_z = 0$ and $q_z = \pi/d$ limits of the Fetter dispersion $\omega_F(q,q_z)$. The white markers indicate the values of ($q,\omega$) at which $\chi''(z,z')$ is plotted in panels c)-f).
\label{fig:chiqomega_simple} }
\end{figure*}

Using the framework outlined in Section \ref{sec:framework}, we calculate the response of an $N$-layer electron gas for $N = 2$ and $N = 20$, employing the Lindhard polarizability in the low-$q$ limit given in Eq. \ref{eq:simplefermigas}. 
We choose $N = 20$ because this approximates the typical number of layers probed in T-EELS measurements on the strange metal Bi-2212.
Figure \ref{fig:chiqomega_simple}(a) shows the resulting inelastic transmission scattering cross section for $N = 2$, which reveals two collective modes corresponding to plasmons from the two layers. One mode appears noticeably weaker than the other; we discuss the origin of this intensity difference below.

The cross-section for $N=20$ is shown in Fig. \ref{fig:chiqomega_simple}(b). The number of modes is significantly larger, leading to the empirical conclusion that the number of plasmon bands is equal to the number of layers. As in the $N=2$ case, half of the bands are fainter. For both the $N=2$ and $N=20$ cases, the different modes are observed to collapse into a single or pair bands at larger $q$.

We now investigate the microscopic origin of these modes by examining the distribution of the charge response in a layer-resolved fashion, i.e., by looking at $-\chi^{\prime\prime}(z,z',q,\omega)$ at fixed $q,\omega$, for different values of $z$ and $z'$. If Figs. \ref{fig:chiqomega_simple}(c-f) we show the value of $-\chi^{\prime\prime}$ for a selection of modes for the $N=20$ case, plotted against the layer indices $z/d$ and $z'/d$. The corresponding values of $q,\omega$ are marked with crosses in Fig. \ref{fig:chiqomega_simple}(b).

For the band marked ``bulk" (Fig. \ref{fig:chiqomega_simple}(c)), $-\chi^{\prime\prime}$ is seen to be distributed among all the layers, with a maximum at the center of the layer stack ($z/d=10$). This suggests that this mode has a bulk character, in which all layers contribute. The charge response of the mode marked ``surface" is shown in panel (d). In contrast to the bulk mode, it shows a response mainly at $z=z'=0$ and $z=z'=20$. This indicates that the response of this mode is localized at the outermost layers, and hence, it is identified as a surface excitation. Panel (e) and (f) shown the charge response of two bands in the fan, labeled as ``Fetter $i$=13" and ``Fetter $i$=16", the index $i$ being counted from the lowest mode. Both modes show a periodic, tiled pattern, but each with a distinct periodicity. This reveals the character of the different modes in the fan: these are standing wave plasmon modes, each with a different periodicity in the $z$ direction. Their dispersion corresponds to the well-known Fetter dispersion for plasmons in an infinite layered system at discrete values of $q_z$ \cite{Fetter1973,Fetter1974}. The upper and lower limits of the Fetter dispersion, at $q_z = 0$ and $q_z = \pi$, respectively, define the bounds of this fan of modes (panel (b), orange lines). 

The reason that half of the modes have a smaller cross-section is explained by considering the symmetry of $-\chi^{\prime\prime}(z,z^{\prime})$. The even-numbered modes correspond to plasmon resonances in which the response is symmetric around $z=z^{\prime}=N/2$, whereas the odd-numbered modes are antisymmetric. Due to momentum conservation in the scattering process, $q_z$ is very small ($<0.01$~\AA$^{-1}$), and the latter terms in equation \ref{eq:layeredcrosssection} get nearly summed out. Similarly, in the $N=2$ case the bright plasmon corresponds to the in-phase mode and the dark plasmon the out-of-phase mode of the two layers.

At larger momenta, i.e., $q > 0.25$, most modes from the Fetter fan collapse onto a single curve, the surface band being the exception. This can be attributed to the magnitude of the Coulomb interaction $V_{2D}(q)$ decreasing with increasing $q$, lowering the coupling between the layers. This means that the character of the dynamic charge response at large $q$ is more two-dimensional, and is likely more sensitive to the properties of the individual layers.

The surface band is influenced in a similar fashion. At $q>0.05$~\AA$^{-1}$, the surface mode separates from the Fetter solution. In the large $q$-limit, $\epsilon_{2D} = 1 - \Pi^0 V \approx 1/(1+\alpha)$, so that there is a plasmon condition 
\begin{equation}
\label{eq:surfplasmon}
\omega_s(q) = \omega_{p,s} \sqrt{q}
\end{equation}
with 
\begin{equation}
 \omega_{p,s} = \sqrt{\frac{2\pi n e^2}{ m_e(\epsilon_0 + \epsilon_1)/2}}.
\end{equation}
In essence, at large $q$, this mode is a two-dimensional plasmon, experiencing an effective dielectric medium with $\epsilon_r = (1 + \epsilon_1)/2$ \cite{Jain1985a}. 

\subsection{Full 2D Lindhard polarizability}
\label{sub:full_2d_lindhard_polarizability}

We now present the cross-section computed using the full two-dimensional Lindhard polarizability, as given in Eq. \ref{eq:lindhard}, which includes quasiparticle excitations. As before, calculations are carried out for $N = 2$ and $N = 20$, with all geometric parameters identical to those used previously. Comparing these results to those in Fig. \ref{fig:chiqomega_simple} helps distinguish which features of the response arise purely from the layered geometry and interlayer Coulomb interaction, and which reflect intrinsic properties of the individual layers.

\begin{figure*}
\includegraphics[width=\textwidth]{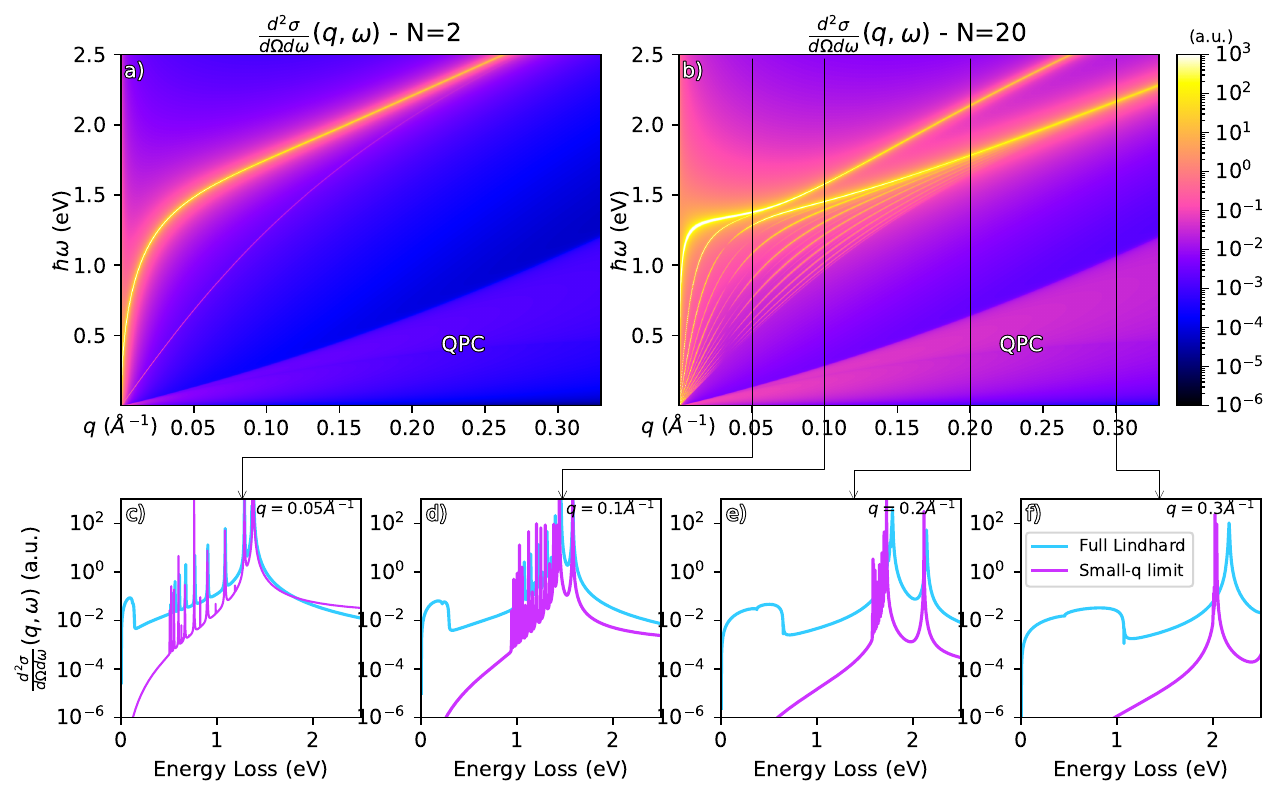}
\caption{The simulated scattering cross section of a T-EELS experiment of (a) an $N=2$ and (b) an $N=20$ layered electron gas using the full Lindhard single-layer polarizability. The quasiparticle continuum is marked in both panels with the text QPC. Panels (c)-(f) show line plots of the cross-section shown in panel (b) at various fixed $q$, indicated by the arrows. The energy of the incident probe electrons is taken to be $E_0$=60~keV.
\label{fig:chi_lindhard} }
\end{figure*}

Figures \ref{fig:chi_lindhard}(a) and (b) show the computed T-EELS cross-sections for $N = 2$ and $N = 20$, respectively, using the full Lindhard polarizability. As expected, both cases closely resemble Fig. \ref{fig:chiqomega_simple} at small $q$. The number of modes and their dispersions are largely unchanged, with the main difference being the appearance of a quasiparticle continuum (QPC) at low energy and finite momentum, labeled in both panels.

For a detailed comparison between the response calculated using the full Lindhard polarizability and that obtained in the small-momentum limit, we show the cross-sections for the $N = 20$ case as 2D line plots at selected momenta in Figs. \ref{fig:chi_lindhard}(c)–(f).
For $q < 0.1$ Å$^{-1}$ the positions and lineshapes of the plasmon excitations are similar between the two cases. 
The main difference is additional spectral weight at energies $E < 1$~eV in the calculation using the full Lindhard polarizability, which is attributed to the presence of the quasiparticle continuum. 

At larger $q$, Figs. \ref{fig:chi_lindhard}(e)–(f), the two cases look quite different. In the calculation using the full Lindhard polarizability, the plasmon modes appear at higher energies and are accompanied by a strong continuum that is absent in the small-momentum-limit case. Since both calculations share the same geometry, this difference can be attributed to the presence of quasiparticle excitations. These effects become more pronounced at larger $q$, where the interlayer Coulomb interaction $V_{2D}(q)$ weakens, causing the overall response to more closely resemble that of a single layer described by $\Pi^0$. This suggests, generically, that at higher momenta the response increasingly reflects the intrinsic, macroscopic properties of the individual layers.

Overall, we conclude that the T-EELS response at small $q$---at least for a layered metal---is relatively generic, governed primarily by the system’s geometry and the long-range nature of the Coulomb interaction. At larger $q$, however, the response exhibits a pronounced single-particle continuum that is highly sensitive to the microscopic details of the individual layers.

\section{Strange metal response from optical data}
We now compute the scattering cross-section of the layered cuprate Bi$_2$Sr$_2$CaCu$_2$O$_{8+x}$ (Bi-2212), a strange metal for which prior EELS measurements have produced conflicting results, particularly at $q \neq 0$. \cite{Nucker1989,Nucker1991,Wang1990,Terauchi1995,Terauchi1999a}. 
Our goal is to determine the expected response of a strange metal and, by comparison with the previously computed Lindhard cases, identify which features are generic to layered metals and which are specific to the strange metal state.

Bi-2212 consists of Cu-O planes sandwiched between charge reservoir layers. 
In our model, the Cu-O bilayers in each unit cell are treated as a single two-dimensional electron gas layer, with the interlayer spacing set by the unit cell dimensions. The reason for doing so is the small spacing between CuO planes in the bilayer, with just a Ca plane in between. At small $q$, explicitly including the bilayer structure would result in a rather small correction, which ignores all possible interaction but Coulomb.

\begin{figure*}
\includegraphics[width=\textwidth]{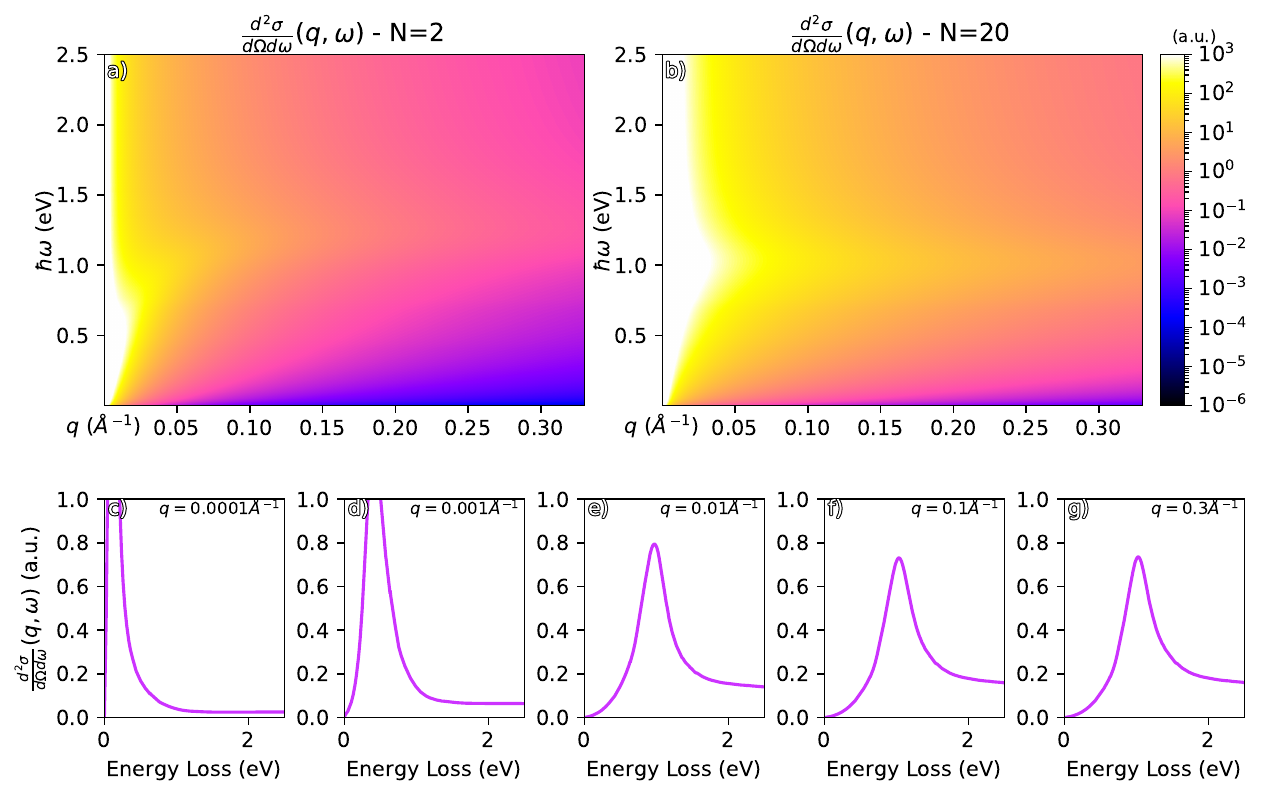}
\caption{The simulated scattering cross section for a T-EELS experiment of (a) an $N=2$ and (b) an $N=20$ layered strange metal. (c)-(g) The data in panel (b) shown as normalized line plots for a selection of momenta. The energy of the incident probe electrons is taken to be $E_0$=60 keV.
\label{fig:chi_bscco} }
\end{figure*}

For our calculations, we use the individual layer polarizability at $q=0$ determined from IR optics experiments, which have been shown to be highly reproducible among multiple groups \cite{Marel2003b,Levallois2016,Basov2005}.
IR optics measures the complex optical conductivity $\sigma(\omega) = \sigma_1(\omega) + i\sigma_2(\omega)$ or, equivalently, the frequency-dependent dielectric function in the optical limit ($q\approx 0$) via
\begin{equation}
\epsilon(\omega) = \epsilon_0 + i \frac{\sigma(\omega)}{\omega}.     
\end{equation}
Assuming that the layers interact only through the long-range Coulomb interaction, the dielectric function can be related to the single-layer polarizability $\Pi^0(q,\omega)$ as \cite{Chen2024}
\begin{equation}
\label{eq:jin_pi}
\Pi^0(q,\omega)\bigr \rvert_{q=0} = \frac{\epsilon_{\infty} - \epsilon(\omega)}{V_{2D}(q)F(q)},
\end{equation}
where $\epsilon_{\infty}$ is the effective background dielectric constant, and $F(q)$ the structure factor for the 3D Coulomb interaction in layered materials \cite{Shung1986,Reed2010}, defined as
\begin{equation}
F(q) = \frac{\sinh(qd)}{\cosh(qd) - 1} .
\end{equation}
We then calculate the inelastic electron scattering cross-section as in the previous sections. The resulting response across a range of $q,\omega$ is shown in Figs. \ref{fig:chi_bscco}(a) and (b) for the $N=2$ and $N=20$ cases, respectively.

The primary feature in both cross-sections is a single peak, which at momentum $q>0.05$~\AA$^{-1}$ is centered at an energy of 1.3~eV in the $N=2$ case and at 1.1~eV in for $N=20$. In contrast to the Lindhard layered electron gas, no Fetter-like fan of discrete plasmon modes is observed, and the quasiparticle continuum characteristic of the Lindhard case is absent.

To examine the response in more detail, Fig. \ref{fig:chi_bscco}(c)–(f) shows the cross-section as a function of energy for the $N=20$ system, plotted as line cuts at selected momenta ($q = 0.0001$, 0.001, 0.01, 0.1, and 0.3 \AA$^{-1}$). These plots are displayed on a linear scale, in contrast to the logarithmic scale used in Fig. \ref{fig:chi_lindhard}.

Several striking differences are observed in the layered strange metal compared to the Lindhard case. First, only a single peak is visible at all momenta, unlike the Lindhard case in which the number of plasmon modes is equal to the number of layers. This includes the surface mode, which is clearly seen in the Lindhard response but is absent in the strange metal. Second, the peak linewidth is significantly broader than in the Lindhard case, indicating a much shorter lifetime. Third, the plasmon peak in the strange metal disperses only for very small momenta ($q \leq 0.01$ \AA$^{-1}$), whereas the Lindhard plasmons exhibit strong dispersion at all $q$. Finally, the quasiparticle continuum characteristic of the Lindhard response is absent in the strange metal.

Based on these these observations, we can now deduce what part of the response is unique to the strange metal.
The broad peak can be interpreted as a highly damped plasmon. Due to the significant linewidth of this peak, which reflects a highly dissipative nature of the collective mode, the individual standing modes expected at small $q$, as well as the surface mode, are blurred together and cannot be resolved. 
The plasmon mode exhibits dispersion for very small momenta $q < 0.01$~\AA$^{-1}$, which is purely the result of the Coulomb interaction between the layers. 
At larger momenta, the response shows little $q$-dependence, consistent with the expectation that it is governed only by the single-layer polarizability $\Pi^0(q\sim 0,\omega)$. 

\section{\label{sec:conclusion}Conclusion}

Understanding the strange metal phase requires knowledge of its 
momentum- and energy-dependent dynamic charge susceptibility, $\chi(q,\omega)$. Electron energy-loss spectroscopy (EELS) probes $\chi(q,\omega)$, yet experiments at finite momentum, $q$, report conflicting results. Clarifying how both transmission (T-EELS) and reflection (R-EELS) cross-sections relate to the bulk susceptibility is essential for progress on the strange metal problem. Here, we semi-empirically calculate the T-EELS charge response of a finite stack of metallic layers, comparing a ``textbook" Lindhard electron gas with a layered strange metal. Our approach in the strange metal case uses the polarizability, $\Pi^0(q,\omega)$, from highly reproducible infrared (IR) optics experiments, assumed to be momentum-independent for modest values of $q$ (i.e., less than the Fermi momentum).

For the layered Lindhard metal, the low-$q$ response is dominated by standing-wave plasmon modes with even and odd parity with respect to the center of the layer, reflecting a lack of $q_z$ conservation in a 20-layer stack. These excitations mainly depend on the geometry of the system and the interlayer Coulomb interaction, with layer-specific details emerging only at large $q$.
The surface plasmon resides at a higher energy than the bulk modes, differing fundamentally from a homogeneous metal in which $\omega_{sp} \approx \omega_p / \sqrt{2}$. In the strange metal, the complex polarizability $\Pi^0$ produces a strongly damped, weakly dispersive plasmon, suppressing Fetter-like and surface plasmon modes. While our calculations based on IR data reproduce T-EELS results at low-$q$, they match no published EELS data at larger $q$. Resolving these discrepancies therefore remains an experimental challenge.

\section{Acknowledgments}
We gratefully acknowledge J. Zaanen, B. Uchoa, D. Ba\l ut, S. Aguirre, and K. Flipse for helpful discussions, and D. van der Marel for supplying infrared data. 
This work was supported by the Center for Quantum Sensing and Quantum Materials, an Energy Frontier Research Center funded by the U.S. Department of Energy (DOE), Office of Science, Basic Energy Sciences
(BES), under Award No. DE-SC0021238. P.A. acknowledges additional support from the EPiQS program of the Gordon and Betty Moore
Foundation under Grant No. GBMF9452. 

\bibliography{bibliography}

\end{document}